\RequirePackage{ifpdf}
\ifpdf 
\documentclass[pdftex]{sigma}
\else
\documentclass{sigma}
\fi

\begin{document}
\allowdisplaybreaks

\renewcommand{\PaperNumber}{064}

\FirstPageHeading

\ShortArticleName{On a `Mysterious' Case of a Quadratic
Hamiltonian}

\ArticleName{On a `Mysterious' Case of a Quadratic Hamiltonian}

\Author{Sergei SAKOVICH}

\AuthorNameForHeading{S. Sakovich}

\Address{Institute of Physics, National Academy of Sciences,
220072 Minsk, Belarus}

\Email{\href{mailto:saks@tut.by}{saks@tut.by}}

\ArticleDates{Received June 02, 2006, in f\/inal form July 18,
2006; Published online July 28, 2006}

\Abstract{We show that one of the f\/ive cases of a quadratic
Hamiltonian, which were recently selected by Sokolov and Wolf who
used the Kovalevskaya--Lyapunov test, fails to pass the
Painlev\'{e} test for integrability.}


\Classification{70H07; 37J30; 34M35}

\section{Introduction}

Recently, Sokolov and Wolf \cite{SW} applied the
Kovalevskaya--Lyapunov test for integrability to a~class of
quadratic Hamiltonians and selected in this way f\/ive cases, out
of which three cases were previously known to be integrable and
one case turned out to be a new integrable Hamiltonian on
$so(3,1)$ with an additional sixth-degree polynomial integral.
Integrability of the f\/ifth case remained unknown, and the
authors of \cite{SW} wrote the following:
\begin{quote}
Case (e) is a mysterious one. We have verif\/ied that the
Hamiltonian has no additional polynomial integrals of degrees less
than or equal to~8. On the other hand, on all Kowalewski solutions
all Kowalewski exponents are integers. It would be interesting to
verify whether the equations of motion in case (e) satisfy the
standard Painlev\'{e} test.
\end{quote}
In the present short note, we apply the Painlev\'{e} test for
integrability to this `mysterious' case (e) and show that it must
be nonintegrable due to some movable logarithmic branching of
solutions. We use the Ablowitz--Ramani--Segur algorithm of
singularity analysis of ODEs \cite{ARS} (see also the review
\cite{RGB}).

\section{The studied quadratic Hamiltonian}

Sokolov and Wolf \cite{SW} considered the following family of
Poisson brackets:
\begin{gather}
    \{ M_i , M_j \} = \varepsilon_{ijk} M_k , \qquad \{ M_i , \gamma_j \}
    = \varepsilon_{ijk} \gamma_k , \qquad \{ \gamma_i , \gamma_j \} = \kappa \varepsilon_{ijk} M_k , \label{pb}
\end{gather}
where $M_i$ and $\gamma_i$ are components of three-dimensional
vectors ${\boldsymbol M}$ and ${\boldsymbol \Gamma}$,
$\varepsilon_{ijk}$ is the totally skew-symmetric tensor, and
$\kappa$ is a parameter. The cases $\kappa > 0$ and $\kappa < 0$
correspond to the $so(4)$ and $so(3,1)$ Lie algebras, whereas the
$e(3)$ Lie algebra case $\kappa = 0$ was not studied in \cite{SW}.
Since the brackets \eqref{pb} have two Casimir functions, $J_1 = (
{\boldsymbol M} , {\boldsymbol \Gamma} )$ and $J_2 = \kappa |
{\boldsymbol M} |^2 + | {\boldsymbol \Gamma} |^2$ (with standard
notations for the vector dot product and module), only one
additional integral functionally independent of the Hamiltonian
and the Casimir functions is required for the Liouville
integrability of the equations of motion.

The special class of Hamiltonians studied by Sokolov and Wolf
\cite{SW} has the following form:
\begin{gather}
    H = c_1 ( {\boldsymbol a} , {\boldsymbol b} ) | {\boldsymbol M} |^2 + c_2 ( {\boldsymbol a} , {\boldsymbol M} ) ( {\boldsymbol b} ,
    {\boldsymbol M} ) + ( {\boldsymbol b} , {\boldsymbol M} \times {\boldsymbol \Gamma}) , \label{qh}
\end{gather}
where $c_1$ and $c_2$ are parameters, the constant vectors
${\boldsymbol a}$ and ${\boldsymbol b}$ are taken to be
${\boldsymbol a} = ( a_1 , 0 , a_3)$ and ${\boldsymbol b} = ( 0 ,
0 , 1)$, $a_1^2 + a_3^2 = - \kappa$, and $\times$ denotes the
vector skew product. The Hamiltonians \eqref{qh}, referred to as
`vectorial' Hamiltonians in \cite{SW}, belong to a wider class of
quadratic Hamiltonians which have numerous applications (two-spin
interactions, motion of a three-dimensional rigid body in a
constant-curvature space or in an ideal f\/luid, motion of a body
with ellipsoidal cavities f\/illed with an ideal f\/luid, etc).

Sokolov and Wolf \cite{SW} applied the Kovalevskaya--Lyapunov test
for integrability to the class of Hamiltonians \eqref{qh} and
selected in this way the following f\/ive cases: (a) $c_1$ is
arbitrary, $c_2 = 0$; (b)~$c_1 = 1$, $c_2 = -2$; (c) $c_1 = 1$,
$c_2 = -1$;  (d) $c_1 = 1$, $c_2 = - \frac{1}{2}$; (e) $c_1 = 1$,
$c_2 = 1$. It was pointed out in \cite{SW} that the case (a)
obviously possesses the linear additional integral $I = (
{\boldsymbol b} , {\boldsymbol M} )$,
 whereas the cases (b) and (c) correspond to two recently discovered integrable
 Hamiltonian systems with polynomial additional integrals of degrees three \cite{TG}
 and four \cite{S}, respectively. It is remarkable that the case (d) turned out to be
 a new integrable quadratic Hamiltonian discovered by the Kovalevskaya--Lyapunov test,
 with an additional sixth-degree polynomial integral \cite{SW}. However,
 in the case (e), which also passed the Kovalevskaya--Lyapunov test well,
 Sokolov and Wolf \cite{SW} failed to f\/ind an additional integral.

In the next section, we will show that the equations of motion in
this `mysterious' case (e) fail to pass the Painlev\'{e} test and
must be nonintegrable in the Liouville sense. It is worthwhile to
remember, however, that the fact of not passing the Painlev\'{e}
test does not necessarily imply nonintegrability of a tested
system, and the way is essential how the test is failed. If the
Painlev\'{e} test is failed at its f\/irst step, when the leading
exponents are determined, or at its second step, when the
positions of resonances are determined, it is sometimes possible
to improve the behavior of solutions by a transformation of
variables so that the transformed system passes the Painlev\'{e}
test well. In particular, rational leading exponents and rational
positions of resonances are allowed by the so-called weak
Painlev\'{e} property which may \cite{RDG}~-- as well as may not
\cite{GDR}~-- correspond to integrability. However, if the
Painlev\'{e} test is failed at its third step, when the
compatibility of recursion relations is checked at the resonances,
one only have to introduce some logarithmic terms into the
expansions of solutions, and this logarithmic branching of
solutions is generally believed to be a clear symptom of
nonintegrability~\cite{AC}. We will see that, in the case~(e) of
the Hamiltonian \eqref{qh}, the Painlev\'{e} test is failed in
this last~-- hopeless~-- way.

\section{Demystifying the `mysterious' case}

We are going to show that the system of six ODEs
\begin{gather}
    \dot{m}_1  = m_1 m_2 + 2 a m_2 m_3 - m_1 g_3 + m_3 g_1 , \nonumber\\
    \dot{m}_2  = - m_1^2 - 2 a m_1 m_3 + m_3^2 - m_2 g_3 + m_3 g_2 , \nonumber\\
    \dot{m}_3  = - m_2 m_3 , \nonumber\\
    \dot{g}_1  = \left( 1 + a^2 \right) m_1 m_3 + m_1 g_2 - 2 a m_2 g_3 + 4 a m_3 g_2 + g_1 g_3 , \nonumber\\
    \dot{g}_2  = \left( 1 + a^2 \right) m_2 m_3 - m_1 g_1 + 2 a m_1 g_3 - 4 a m_3 g_1 + m_3 g_3 + g_2 g_3 ,\nonumber\\
    \dot{g}_3  = - \left( 1 + a^2 \right) \left( m_1^2 + m_2^2 \right) - 2 a m_1 g_2 + 2 a m_2 g_1 - m_3 g_2 - g_1^2 - g_2^2 ,
\label{em}
\end{gather}
where the dot denotes $\frac{d}{dt}$ and $a$ is a parameter,
 must be nonintegrable unless $a = 0$. The system~\eqref{em}
 represents the equations of motion in the case (e) of the Hamiltonian \eqref{qh}.
 The notations $m_i$, $g_i$ and $a$ used in \eqref{em} correspond to $M_i$, $\gamma_i$ and $a_3$
 used in \eqref{qh}, respectively, whereas for the nonzero parameter $a_1$
 of \eqref{qh} we have set $a_1 = 1$ by rescaling $M_i$ and $\gamma_i$
 (note that the case (e) with $a_1 = 0$ falls under the case (a) which possesses a linear additional integral).

Assuming that $a \neq 0$ in \eqref{em}, where all quantities are
considered as complex-valued from now on, and using the expansions
\begin{gather}
    m_i  = m_{i,0} \phi^{\alpha_i} + \dotsb + m_{i,r} \phi^{r + \alpha_i} + \dotsb ,\nonumber \\
    g_i  = g_{i,0} \phi^{\beta_i} + \dotsb + g_{i,r} \phi^{r + \beta_i} + \dotsb , \nonumber\\
    i  = 1, 2, 3, \qquad \phi = t - t_0 ,
\label{ex}
\end{gather}
where $m_{i,j}$, $g_{i,j}$, $\alpha_i$, $\beta_i$ and $t_0$ are
constants, we f\/ind the following singular branches and
positions~$r$ of resonances in them (note that we do not consider
the possibility of $a^2 = -1$, because it corresponds to the case
of $\kappa = 0$, i.e.\ the $e(3)$ Lie algebra, which was not
studied in \cite{SW}):
\begin{gather}
    \alpha_i  = \beta_i =-1, \qquad i = 1, 2, 3, \nonumber\\
    m_{1,0}  = \frac{\pm a}{\sqrt{-1-a^2}} , \qquad m_{2,0} = 1 , \qquad m_{3,0} =
\frac{\mp 1}{\sqrt{-1-a^2}} , \nonumber\\
    g_{1,0}  = - a , \qquad g_{2,0} = \mp \sqrt{-1-a^2} , \qquad g_{3,0} = 1 , \nonumber\\
    r  = - 1, 0, 1, 2, 2, 2;
\label{b1}
\end{gather}
and
\begin{gather}
    \alpha_i  = \beta_i =-1, \qquad i = 1, 2, 3, \nonumber\\
    m_{1,0}  = \frac{\pm a}{\sqrt{-1-a^2}} , \qquad m_{2,0} = 1 , \qquad m_{3,0} =
\frac{\mp 1}{\sqrt{-1-a^2}} , \nonumber\\
    g_{1,0}  = a , \qquad g_{2,0} = \pm \sqrt{-1-a^2} , \qquad g_{3,0} = - 1 , \nonumber\\
    r  = - 2, - 1, 1, 2, 2, 4.
\label{b2}
\end{gather}

Let us look at the branch \eqref{b1} f\/irst. According to the
positions of resonances, this branch must be a generic one. We
see, however, that the position of one resonance is $r=0$ there,
whereas all the coef\/f\/icients $m_{1,0}$, $m_{2,0}$, $m_{3,0}$,
$g_{1,0}$, $g_{2,0}$ and $g_{3,0}$ turn out to be f\/ixed. This
means that, in the case of \eqref{b1}, the recursion relations for
the coef\/f\/icients of the expansions \eqref{ex} have a
nontrivial compatibility condition right in the position $r=0$,
and we have to modify \eqref{ex} by introducing additional
logarithmic terms, starting from the terms proportional to
$\phi^{-1} \log \phi$. Suppose we do not do this and think that
the branch \eqref{b1} represents not the general solution but a
class of special solutions. Then we f\/ind that no compatibility
condition appears at the resonance $r=1$, where the
coef\/f\/icient $m_{1,1}$ remains arbitrary. However, at the
triple resonance $r=2$, where the coef\/f\/icients $m_{1,2}$,
$m_{2,2}$ and $g_{1,2}$ remain arbitrary, the nontrivial
compatibility condition $m_{1,1}=0$ appears, and we are again
forced to introduce logarithmic terms into the expansions of
solutions.

The branch \eqref{b2} gives us the same information: solutions of
the system \eqref{em} with $a \neq 0$ possess movable logarithmic
singularities. In this branch, the nontrivial compatibility
condition
\begin{gather}
     m_{1,1}^2 \left[ 2 a \sqrt{-1-a^2} \left( 2+a^2 \right)^2 m_{1,2} \right. \notag \\
     \qquad \qquad \left. + \left( 2+a^2 \right)^2 \left( 3+2a^2 \right) m_{2,2}
    + \left( 4+9a^2+5a^4 \right) m_{1,1}^2 \right] = 0
\end{gather}
appears at the resonance $r=4$.

Consequently, the system \eqref{em} with $a \neq 0$ fails to pass
the Painlev\'{e} test for integrability, and this, in its turn,
explains why Sokolov and Wolf \cite{SW} failed to f\/ind an
additional integral for the case~(e) $c_1 = 1$, $c_2 = 1$ of the
quadratic Hamiltonian \eqref{qh}.

The case of \eqref{em} with $a=0$ is dif\/ferent: it passes the
Painlev\'{e} test well, as one can easily verify. However, the
fact of integrability of the system \eqref{em} with $a=0$ is not
new: the case~(e) with $a_3 = 0$ of the Hamiltonian \eqref{qh}
falls under the case~(c) with a fourth-degree additional
polynomial integral (we are indebted to
 Prof.~T.~Wolf who kindly explained this point to us).

\section{Conclusion}

We have shown that the `mysterious' case (e) $c_1 = 1$, $c_2 = 1$
of the quadratic Hamiltonian \eqref{qh}, selected by Sokolov and
Wolf \cite{SW} who used the Kovalevskaya--Lyapunov test, fails to
pass the Painlev\'{e} test for integrability. It is worthwhile to
remember that, from the standpoint of the Painlev\'{e} test
\cite{ARS,RGB}, the Kovalevskaya--Lyapunov test only examines the
positions of resonances for a studied system and is sensitive to
the nondominant algebraic branching of solutions, whereas the
Painlev\'{e} test also verif\/ies the compatibility of recursion
relations at the resonances and can detect the nondominant
logarithmic branching of solutions which is the strongest
indication of nonintegrability. We can add that it happens quite
frequently in the singularity analysis practice that the positions
of resonances are good but the recursion relations are
incompatible. For example, in the integrability study of
symmetrically coupled higher-order nonlinear Schr\"{o}dinger
equations \cite{ST1}, 23 distinct cases with integer positions of
all resonances were found, but the recursion relations turned out
to be compatible in only one case out of those 23 cases (see the
table in \cite{ST2} for more details).

\LastPageEnding

\end{document}